\documentclass[prd,aps,floats,floatfix,eqsecnum,nofootinbib]{revtex4}
\newcommand{\lsim}{\lower 2pt \hbox{$\, \buildrel {\scriptstyle
<}\over {\scriptstyle \sim}\,$}}  \newcommand{\gsim}{\lower 2pt
\hbox{$\, \buildrel {\scriptstyle >}\over {\scriptstyle \sim}\,$}}
\usepackage{graphics}
\usepackage{graphicx}
\usepackage{rotate}
\usepackage{color}
\usepackage{rotating}
\usepackage{psfrag}
\usepackage{amsmath,amsthm}
\usepackage{epsfig}
\usepackage{slashed}
\usepackage{sidecap}

\begin{document}

\setcounter{secnumdepth}{1}

\newcommand{\D}{\displaystyle} 
\newcommand{\T}{\textstyle}        
\newcommand{\SC}{\scriptstyle}  
\newcommand{\SSC}{\scriptscriptstyle} 
\newcommand{\be}{\begin{equation}}
\newcommand{\ee}{\end{equation}}
\newcommand{\avg}[1]{\langle #1 \rangle}
\newcommand{\vx}{{\boldsymbol{x}}}
\newcommand{\rv}{{\boldsymbol{r}}}
\newcommand{\vq}{\ensuremath{\vec{q}}}
\newcommand{\pv}{\ensuremath{\vec{p}}}
\hyphenation{}

\title{\Large     

Summaries and Conclusions of the workshop:

`Extreme Starbursts in the Local Universe'



Instituto de Astrof\'\i sica de Andaluc\'\i a (CSIC)

Granada, 21-25 June 2010}

\author{ \bf   R.M. Gonz\'alez Delgado$^{(1)}$,   
J. Rodr\'\i guez Zaur\'\i n$^{(2)}$, 
E. P\'erez$^{(1)}$,
A. Alonso-Herrero $^{(2)}$, 
C. Tadhunter $^{(3)}$,   
S. Veilleux$^{(4)}$, 
T. Heckman$^{(5)}$, 
R. Overzier$^{(6)}$, 
T.S. Gon\c{c}alves$^{(7)}$, 
A. Alberdi$^{(1)}$,
M.A. P\'erez Torres$^{(1)}$, 
A. Pasquali$^{(8)}$, 
A. Monreal-Ibero$^{(1)}$, 
T. D\'\i az-Santos$^{(9)}$, 
S. Garc\'\i a-Burillo$^{(10)}$, 
D. Miralles Caballero$^{(2)}$, 
P.  Di Matteo$^{(11)}$, 
L. Kewley$^{(12)}$, 
C. Ramos Almeida$^{(3)}$, 
B. Weiner$^{(13)}$, 
B. Rothberg$^{(14)}$, 
J.C. Tan$^{(15)}$, 
S. Jogee$^{(16)}$, 
R. Cid Fernandes$^{(17)}$, 
M. Rodrigues$^{(18)}$, 
R. Delgado-Serrano$^{(18)}$, 
H. Spoon$^{(19)}$, 
P. Hopkins$^{(20)}$, 
D.  Rupke$^{(13)}$, 
E. Bellocchi$^{(2)}$, 
C. Cortijo$^{(1)}$,
J. Piqueras L\'opez$^{(2)}$, 
G. Canalizo$^{(21)}$, 
M. Imanishi$^{(22)}$, 
M. Lazarova$^{(21)}$, 
M. Villar-Mart\'\i n$^{(1)}$,
 M. Brotherton$^{(23)}$, 
 V. Wild$^{(24)}$, 
 M. Swinbank$^{(25)}$, 
 K. Menendez-Delmestre$^{(26)}$, 
 F. Hammer$^{(18)}$, 
 P. P\'erez-Gonz\'alez$^{(27)}$, 
 J. Turner$^{(28)}$, 
 J. Fischer$^{(15)}$, 
 S. F. S\'anchez$^{(29)}$, 
 L. Colina$^{(2)}$, 
 A. Gardini$^{(1)}$ }

\date{\today}

\affiliation{
$^{(1)}$ Instituto de Astrof\'\i sica de Andaluc\'\i a CSIC (Spain);
$^{(2)}$ Centro de Astrobiolog\'\i a CSIC (Spain);
$^{(3)}$ University of Sheffield (UK);
$^{(4)}$ University of Maryland (USA);
$^{(5)}$ Johns Hopkins University (USA);
$^{(6)}$ Max-Planck-Institut f\"ur Astrophysics (Germany);
$^{(7)}$ Caltech (USA);
$^{(8)}$ Max-Planck-Institut f\"ur Astrophysics (Germany);
$^{(9)}$ University of Crete;
$^{(10)}$ Observatorio Astron\'omico Nacional (Spain);
$^{(11)}$ Observatoire de Paris (France);
$^{(12)}$ Institute for Astronomy, University of Hawaii (USA);
$^{(13)}$ Steward Observatory (USA);
$^{(14)}$ Naval Research Laboratory (USA);
$^{(15)}$ University of Florida (USA);
$^{(16)}$ University of Texas at Austin (USA);
$^{(17)}$ Universidad Federal de Santa Catarina (Brasil);
$^{(18)}$ GEPI- Observatoire de Paris (France);
$^{(19)}$ Cornell University (USA);
$^{(20)}$ University of California, Berkeley (USA);
$^{(21)}$ University of California, Riverside (USA);
$^{(22)}$ National Astronomical Observatory of Japan;
$^{(23)}$ University of Wyoming (USA);
$^{(24)}$ Institut d'Astrophysique de Paris (France);
$^{(25)}$ Durham University (UK);
$^{(26)}$ Carnegie Observatoires (USA);
$^{(27)}$ Universidad Complutense de Madrid (Spain);
$^{(28)}$ UCLA (USA);
$^{(29)}$ Centro Astron\'omico Hispano Alem\'an (CAHA CSIC, Spain)
}

\begin{figure}[http*]
\includegraphics[scale=0.4]{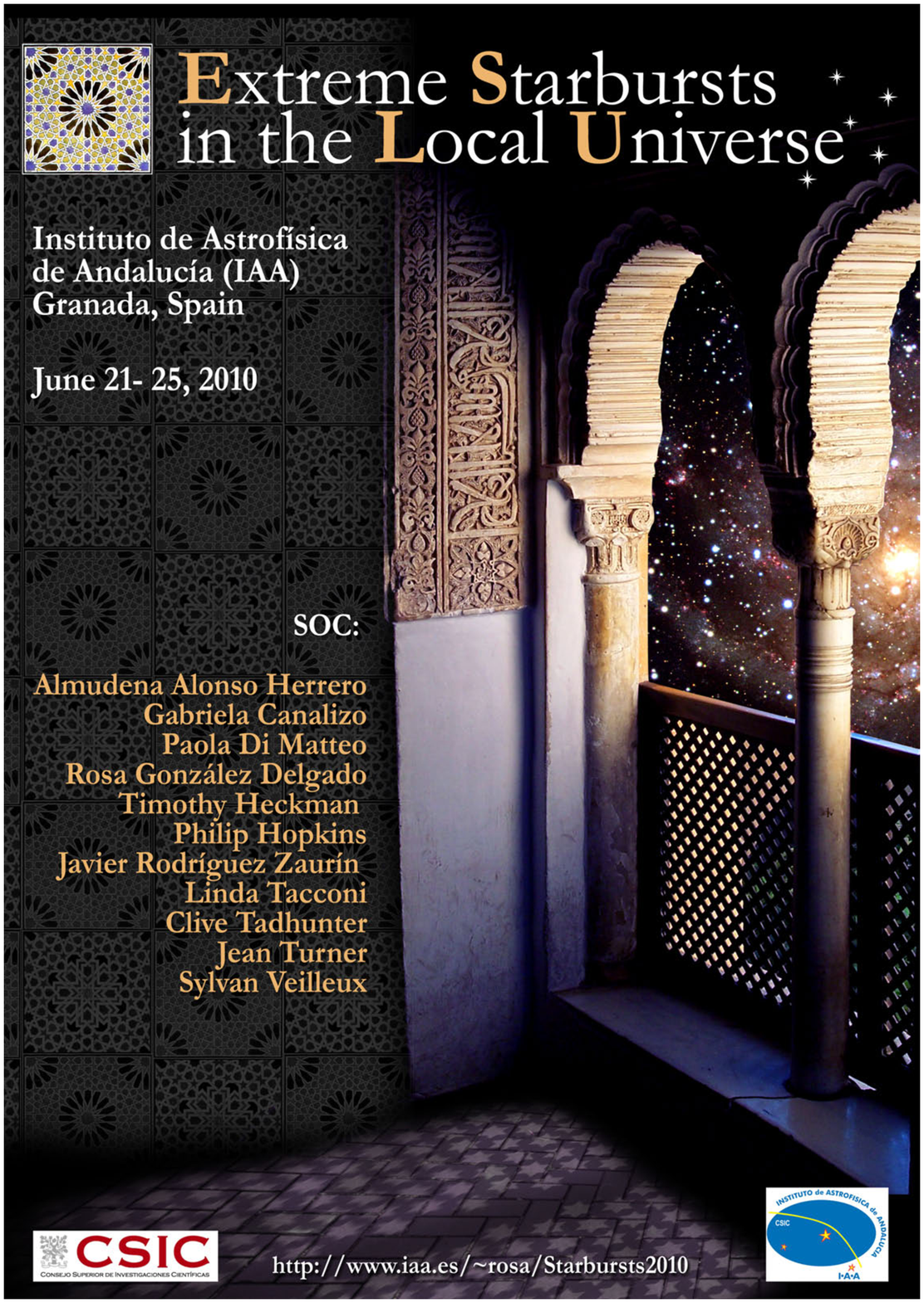}
\end{figure}

\maketitle



\newpage

\section{Purpose of the Workshop and Introduction}

Despite their general importance for understanding the evolution of galaxies at all redshifts, there remain considerable uncertainties about the physical mechanisms involved in triggering starbursts, and also how such events relate to the general properties of the host galaxies and their nuclei. Over the last decade there has been considerable observational progress in studying starbursts with the current generation of ground- and space-based facilities, particularly in uncovering heavily dust-obscured regions, studying the stellar populations in the host galaxies, and pushing the envelope of studies of starbursts to ever higher redshifts; this rate of progress is set to increase even further with the new generation of facilities such as JWST, Herschel, ALMA, eVLA, eMerlin, etc. At the same time, it has become possible to undertake detailed hydrodynamical simulations of star formation in galaxies as they evolve under various circumstances. Therefore the aim of the workshop was to bring together observers and theorists to review the latest results. The purpose of the workshop was to address the following issues: 

\begin{itemize}

\item{What are the main modes of triggering extreme starbursts in the local Universe?}
\item{How efficiently are stars formed in extreme starbursts?}
\item{What are the star formation histories of local starburst galaxies?}
\item{How well do the theoretical simulations model the observations?}
\item{What can we learn about starbursts in the distant Universe through studies of their local counterparts?}
\item{How important is the role of extreme starbursts in the hierarchical assembly of galaxies?}
\item{How are extreme starbursts related to the triggering of AGN in the nuclei of galaxies?}

\end{itemize}

The workshop lasted for five working days at the Instituto de Astrof\'\i sica  de Andaluc\'\i a, that belongs to the Spanish
Research Council CSIC, located in the beautiful city of Granada (Spain). Seven sessions were dedicated to: 

\begin{itemize}

\item{{\bf Session 1:} Cosmological context and observational properties}
\item{{\bf Session 2:} Triggering mechanisms and environments}
\item{{\bf Session 3:} Star formation histories}
\item{{\bf Session 4:} Feedback effects}
\item{{\bf Session 5:} The AGN/Starburst connection}
\item{{\bf Session 6:} The distant universe}
\item{{\bf Session 7:} Summary and future prospects}

\end{itemize}

About 60 persons mainly from Europe, USA, Brasil and Japan participate in the workshop. All the information about the workshop is displayed at:

\begin{center}

{\bf http://Starbursts2010.iaa.es}

\end{center}

The presentations by the participants are available on line (in .pdf format) under `Program' in the link above.

The workshop was designed with emphasis on discussions. For this reason, in addition to the plenary discussions related to the 6 reviews, 10 invited talks and 29 oral contributions, we had six hours  more for discussion on the following topics:

\begin{itemize}

\item{{\bf Discussion 1:} The relevance of obscuration in the stellar content and the efficiency of star formation in starbursts. Tracers of star formation and star formation laws.}
\item{{\bf Discussion 2:} Star formation: Triggering mechanisms. Prescriptions for modeling star formation rate.}
\item{{\bf Discussion 3:} The different techniques to investigate the star formation histories of starbursts: advantages and limitations.}
\item{{\bf Discussion 4:} The different modes of feedback and their impact on the star formation activity.}
\item{{\bf Discussion 5:} How does the AGN affect the surrounding star formation activity? The accuracy of the different diagnostics.}
\item{{\bf Discussion 6:} Are local Starbursts analogs to high-z star forming galaxies?}

\end{itemize}

We would like to thank  Jean Turner, Jonathan Tan, Paola Di Matteo, Philip Hopkins, Roberto Cid Fernandes, Gabriela Canalizo, David Rupke, Vivienne Wild, Sylvain Veilleux and Timothy Heckman for leading and triggering these stimulating discussions. We also thank, specifically, Sylvain Veilleux, Luis Colina, and Jacky Fischer for keeping the speakers to the schedule during the sessions they chaired,  and specially Sylvain Veilleux who managed to synthesize and summarize the workshop.

\vskip0.2cm

It is a pleasure for us to thank  the other members of the Scientific Organizing Committee for their contributions:   Gabriela Canalizo, Paola Di Matteo, Timothy Heckman, Philip Hopkins, Jean Turner, and Sylvain Veilleux. We express our gratitude to the other members of the Local Organizing Committee: Clara Cortijo-Ferrero and Angela Gardini. Very special thanks are for Andr\'es Alonso Herrero  for providing us with a very beautiful  workshop poster announcement  (the JPG file is available in the workshop web page) that was inspired on one of the most famous corners of the Alhambra named ``Balc\'on de la Reina".

\vskip0.2cm

Finally, we thank the Consejo Superior de Investigaciones Cient\'\i ficas (CSIC) and Instituto de Astrof\'\i sica de Andaluc\'\i a
for the financial support and the logistics assistance for the successful organization of this Workshop.

\begin{center}
                                   
With kind regards,

\bigskip  
                                            
Rosa M. Gonz\'alez Delgado,  Javier Rodr\'\i guez Zaur\'\i n, Enrique P\'erez, Almudena Alonso Herrero, \& Clive Tadhunter

\end{center}

\begin{figure}[http*]
\includegraphics[scale=0.6,angle=-90]{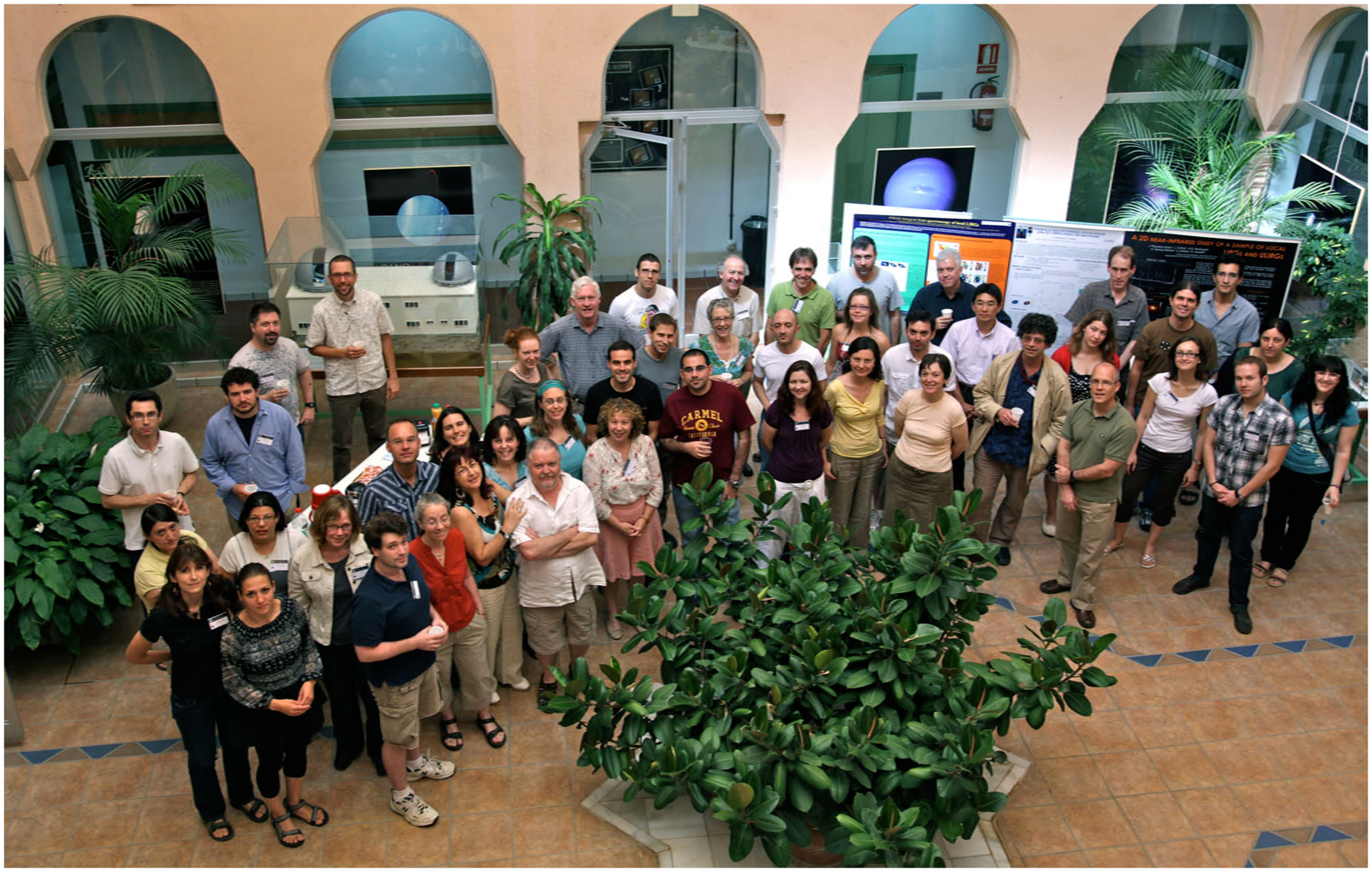}
\end{figure}

\newpage

\section{Summary by the speakers}

Here we present the summary of the speaker«s talk  listed by the order in the program schedule.
The full (in pdf format) contribution can be found at:

\vskip2mm

\centerline{http://Starbursts2010.iaa.es/Starbursts2010/Program.html}

\subsection{\bf Session 1: Cosmological context and observational properties}

{\bf 1.1 T. Heckman}{\it (John Hopkins University, USA)}

\vskip0.2cm

{\it Local Starbursts in a Cosmological Context}

\vskip0.2cm

I will introduce some of the major issues that motivate the conference, with an emphasis on how starbursts fit into the bigger picture. I will argue that local starbursts are unique laboratories in which to study the processes at work in the early Universe. I will define starbursts in several different ways, and discuss the merits and limitations of these definitions. I will argue that the most physically useful definition of a starburst is its intensity (star-formation rate per unit area). This is the most natural parameter to use for comparing local starbursts with physically similar galaxies at high redshift. I will describe how the systematic properties of local starbursts can be understood on the basis of the Schmidt-Kennicutt Law and mass-metallicity relation. I will briefly summarize the properties of starburst-driven galactic superwinds and their possible implications for the evolution of galaxies and the intergalactic medium. These complex multiphase flows are best studied in nearby starbursts, where we can study the hot X-ray gas that contains the bulk of the energy as well as newly produced metals. I will discuss what we can learn from local starbursts about the processes responsible for the reionization of the universe at z $>$ 6. Finally, I will summarize the link between post-starbursts and the growth of supermassive black holes, and will suggest that the lag in black hole growth is caused by feedback from supernovae during the starburst phase.

\bigskip

{\bf 1.2 R. Overzier}{\it (MPIA, Germany)}

\vskip0.2cm

{\it Extreme Starburst Galaxies as Nearby Analogs of High Redshift Lyman Break Galaxies}

\vskip0.2cm

Starburst galaxies are important for our understanding of galaxy evolution at all redshifts. I will present and discuss the latest results from our ongoing survey of ``Local Analogs of Lyman Break Galaxies". Because these starbursts are similar to typical UV-selected starbursts at high redshift in most of their observed and physical properties, we have an excellent training set for understanding the relation between massive star formation, ISM, host galaxy structures, and nuclei of starbursts. In this talk, I will highlight three of our most recent results: 
(1) The nearby sample shows a deviation from the so-called IRX-beta relation that is widely used to infer the ratio of total IR-to-UV luminosities at high redshift. This offset is similar to that found for a few lensed LBGs that have direct detections in the IR. I will show how an improved understanding of the IRX-beta relation directly affects estimates of the cosmic star formation rate history.
(2) The LBG analogs at low redshift and LBGs at high redshift display a range in structures from compact to clumpy that is different from typical local star-forming galaxies. Recent studies have suggested that, at least at high redshift, intense star formation is triggered by massive gas accretion in the form of cold flows. Based on a detailed comparison with the morphologies of LBGs in the Hubble Ultra Deep Field, we conclude, however, that starbursts triggered by mergers remain a viable mechanism for driving the evolution of these starbursts.
(3) Some of the local starbursts display peculiar nuclei that are more massive and more dense than any central star cluster observed to date. We speculate that they are progenitors of the central cusps in low-mass ellipticals being formed in dissipative mergers. The massive, dense nuclei provide an ideal environment for the formation of black holes. New radio and X-ray data suggest the presence of 10$^5$-10$^6$ M$_{\odot}$ black holes.

\bigskip

{\bf 1.3 T.S. Gon\c{c}alves}{\it (Caltech, USA)}

\vskip0.2cm

{\it Integral-Field Observations of Lyman Break Analogs: Lessons for the High-Redshift Universe}

\vskip0.2cm

Lyman Break Analogs (LBAs), characterized by high far-UV luminosities and surface brightnesses as detected by GALEX, are intensely star-forming galaxies in the low-redshift universe (z$\sim$0.2), with star formation rates reaching up to 50 times that of the Milky Way. These objects present metallicities, morphologies and other physical properties similar to higher redshift Lyman-break galaxies (LBGs), motivating the detailed study of LBAs as potential local analogs to this high-redshift galaxy population. Strong hydrogen emission lines and compact sizes make LBAs ideal candidates for integral-field spectrography. We present results for 20 LBAs observed with Keck/OSIRIS with an AO-assisted spatial resolution down to $\sim$200pc. We detect satellite companions, diffuse emission and velocity shear, all with high signal-to-noise ratios. We artificially redshift our data to z=2.2 to allow for a direct comparison with observations of high-z LBGs and find striking similarities between both samples. This suggests that the same physical processes may be responsible for the observed properties of star forming galaxies in the early universe. We also find a strong correlation of observable properties with stellar mass, in which more massive galaxies show greater resemblance to a disk-like structure, not unlike what is observed at high redshifts. Finally, we argue that kinematic studies of the ionized gas in star-forming galaxies is not an appropriate diagnostic to rule out major merger events as the trigger for the observed starburst.

\bigskip

{\bf 1.4 A. Alberdi}{\it (IAA CSIC, Spain)}

\vskip0.2cm

{\it Radio Supernovae: A Window into the Heart of Starburst Galaxies}

\vskip0.2cm

High-resolution radio observations of the nuclear regions of Luminous and Ultraluminous Infrared Galaxies (ULIRGs) have shown that their radio structure consists of a compact high surface-brightness central radio source immersed in a diffuse low brightness circumnuclear halo. While the central components could be associated with AGNs or compact star-forming regions, where radio supernovae are exploding, it is well known that the circumnuclear regions host bursts of star-formation. Studies of radio supernovae can provide essential information about stellar evolution and CSM/ISM properties in regions hidden by dust at optical and IR wavelengths. High-resolution radio observations of LIRGs can allow us to determine the core-collapse supernova rate in them as well as their star-formation rate.
\bigskip

{\bf 1.5 M.A. P\'erez Torres}{\it (IAA CSIC, Spain)}

\vskip0.2cm

{\it An extremely prolific SN factory in the buried nucleus of Arp 299A}

\vskip0.2cm

The central kiloparsec of many local luminous infrared galaxies are known to host intense bursts of massive star formation, leading to numerous explosions of core-collapse supernovae. However, the dust-enshrouded regions where those supernovae explode hamper their detection at optical and near-infrared wavelengths.
We investigated the nuclear region of the starburst galaxy IC 694 (=Arp 299-A) at radio wavelengths, aimed at discovering recently exploded core-collapse supernovae, as well as determining their rate of explosion, which carries crucial information about star formation rates, the initial mass function, and the starburst processes in action.
We used the electronic European VLBI Network (eEVN) to image with milliarcsecond resolution the 5.0 GHz compact radio emission of the innermost nuclear region of IC 694. Our observations detected a rich cluster of 26 compact radio emitting sources in the central 150 pc of the nuclear starburst in IC 694. The high brightness temperatures observed for the compact sources are indicative of a non-thermal origin for the observed radio emission, implying that most, if not all, of those sources are young radio supernovae and supernova remnants. We found evidence of at least three relatively young, slowly evolving, long-lasting radio supernovae that appear to have unusual CCSN properties, suggesting that the conditions in the local circumstellar medium (CSM) play a significant role in determining the radio behavior of expanding SNe. Their radio luminosities are typical of normal RSNe, which result from the explosion of type IIP/b and type IIL SNe. All of these results provide support for a recent (less than 10-15 Myr) instantaneous starburst in the innermost regions of IC 694, and confirm that the inner regions of Arp 299-A are an extremely prolific supernova factory.
\bigskip

{\bf 1.6 A. Pasquali}{\it (MPIA, Germany)}

{\bf A. Bik, S. Zibetti, N. Ageorges, W. Seifert, W. Bradner, H.-W. Rix}

\vskip0.2cm

{\it Exploring local Starbursts with LBT/LUCIFER: the case of NGC 1569}

\vskip0.2cm

We observed NGC 1569 during the commissioning run of LUCIFER, the NIR imager and spectrograph now available at the Large Binocular Telescope. The imaging was performed under good seeing conditions (0.4 arcsec on average) and with the highest angular resolution offered by LUCIFER (0.12 arcsec/pix). We obtained high-quality spatial maps of HeI 1.08, [FeII] 1.64 and Br$\gamma$ emission across the galaxy, and used them together with HST/ACS images in H$\alpha$ and in the V band to derive the 2D map ($1.6\times0.9$ kpc in size) of the dust extinction and surface star-formation rate density across NGC 1569. We show that dust extinction (as derived from the Br$\gamma$/H$\alpha$ flux ratio) is rather patchy and, on average, is higher in the NW portion of the galaxy (E(B-V) = 0.5 mag) than in the SE (E(B-V) = 0.4 mag). Similarly, the surface star-formation rate density peaks in the NW region of NGC 1569, reaching a value of about $3\times10^{-6}$ M$_{\odot}$/yr/kpc$^{2}$. The total star-formation rate as estimated from the integrated, dereddened H$\alpha$ (Br$\gamma$) luminosity is about 0.4 M$_{\odot}$/yr, while the total SN rate from the integrated, dereddened [FeII] luminosity is about 0.005 yr$^{-1}$ (assuming a distance of 3.36 Mpc). The azimuthally averaged [FeII]/Br$\gamma$ flux ratio peaks at the edges of the central cavity (encompassing the super star clusters A and B) and at the galaxy rim. If this line ratio were to compare an older star-formation history (as traced by supernovae) to an on-going activity (represented by OB stars able to ionize hydrogen), it would then indicate that star formation has been quenched within the cavity and is presently occurring in a ring around the cavity.
\bigskip

{\bf 1.7 A. Monreal-Ibero}{\it (ESO and IAA CSIC, Spain)}

{\bf S. Arribas, L. Colina, J. Rodr\'\i guez-Zaur\'\i n, A. Alonso-Herrero, M. Garc\'\i a-Mar\'\i n}
\vskip0.2cm

{\it Ionization mechanisms in the extra-nuclear regions of LIRGs}

\vskip0.2cm

Luminous Infrared Galaxies (LIRGs) are an important class of objects in the low-z universe, bridging the gap between normal spirals and the Ultraluminous Infrared Galaxies. LIRGs are also relevant in a high-z context as a large fraction of the stars in the Universe were formed in these objects. Here, we present a study aiming to understand the nature and origin of the ionization mechanisms operating in the extranuclear regions of LIRGs as a function of the interaction phase and infrared luminosity. The study is based on Integral Field Spectroscopy data obtained with the VIMOS instrument of 32 LIRGs covering all types of morphologies (isolated galaxies, interacting pairs, and advanced mergers), and the entire 10$^{11}$-10$^{12}$ L$_{\odot}$ infrared luminosity range. We found strong evidence for shock ionization, with a clear trend with the dynamical status of the system. Specifically, we quantified the variation with interaction phase of several line ratios indicative of the excitation degree. In particular, while the [NII]/H$\alpha$ ratio does not show any significant change, the [SII]/H$\alpha$ and [OI]/H$\alpha$ ratios are higher for more advanced interaction stages. Also, we constrained the main mechanisms causing the ionization in the extranuclear regions using diagnostic diagrams. Isolated systems are mainly consistent with ionization caused by young stars while large fractions of the extra-nuclear regions in interacting pairs and more advanced mergers are consistent with ionization caused by shocks. The relation between the excitation degree and the velocity dispersion of the ionized gas supports this result. We interpret both results as evidence for shock ionization in interacting galaxies and advanced mergers but not in isolated galaxies. A deeper discussion about these results as well as about the dependence of the velocity dispersion Ð excitation relation with the luminosity of the system can be
found in Monreal-Ibero et al. 2010, A\&A, in press (arXiv:1004.3933).
\bigskip

{\bf 1.8 A. Alonso-Herrero}{\it (CAB CSIC, Spain)}

\vskip0.2cm

{\it The Spitzer view of starburst galaxies}

\vskip0.2cm

In this talk I review some of the new findings for nearby starburst galaxies obtained with the imaging (IRAC and MIPS) and spectroscopy (IRS) instruments on board of the Spitzer SpaceTelescope. In particular, I summarize new results about the most prominent features in the mid-infrared spectra detected in nearby starbursts, such as PAH features, fine structure lines, the 9.7$\mu$m silicate feature, and molecular hydrogen lines. Finally I discuss our progress on using mid-infrared features (e.g., monochromatic luminosities, PAH features, Ne emission lines) as tracers of the star formation rate of galaxies in the local Universe and at high redshift.
\bigskip

{\bf 1.9 T. D\'\i az-Santos}{\it (University of Crete, Crete)}

\vskip0.2cm

{\it The Spatial Extent of (U)LIRGs in the mid-Infrared}

\vskip0.2cm

We present our analysis of the extended mid-Infrared (IR) emission of the local Luminous and Ultraluminous Infrared Galaxies (U)LIRGs comprised in the Great Observatory All-sky LIRG Survey (GOALS) sample. We use Spitzer IRS spectra to determine the fraction of extra-nuclear emission in these (U)LIRGs as a function of wavelength which allow us to compare among different spectral features. We find that in more than 30$\%$ of LIRGs, at least $\sim$50$\%$ of their mid-IR emission stems from the extended component, which may contribute even up to $\sim$80$\%$. As a whole, the mid-IR emission of local LIRGs is 2.5-3 times more extended than that of ULIRGs, suggesting that mid-IR emission of LIRGs is rather distributed across their disks on a scale of several kpc. We find that the compactness of the mid-IR continuum emission at 13.2$\mu$m of the LIRGs/ULIRGs in our sample does not depend on their merging stage, but in turn it is related to the AGN contribution to the mid-IR emission. The more AGN-dominated a system is the less extended appears in the mid-IR. The compactness of the LIRGs/ULIRGs is related with the IRAS 60/100$\mu$m color. Colder galaxies are more extended. This places a lower limit, depending on the far-IR color, to the physical size of the region responsible for the fir-IR continuum emission of these systems, which will soon to be probed by Herschel.
\bigskip

{\bf 1.10 S. Garc\'\i a-Burillo}{\it (OAN, Spain)}

\vskip0.2cm

{\it Star formation laws in LIRGs/ULIRGs}

\vskip0.2cm

We have used the IRAM 30m telescope to observe a sample of 15 LIRGs simultaneously in the 1--0 lines of HCN and HCO$^+$. With the proposed observations we have significantly improved the statistics of LIRGs where high-quality data are available for these key molecular probes of the dense gas content. These observations complement the survey of LIRGs and ULIRGs made by Gracia-Carpio et al. (2006, 2008) and make it possible to build  a final sample of 24 LIRGs with HCN and HCO$^+$ data. Both the star formation rates (SFR) and the typical sizes of the star forming (SF) regions of the galaxies in our sample are well characterized through available high-resolution imaging at different wavelengths (Alonso-Herrero et al 2006). We analyze the star formation efficiency and the SF law derived for the dense molecular gas as traced by HCN(1--0) and HCO$^+$(1-0) in 24 LIRGs. Results issued from these observations will be discussed in the context of the currently debated SF laws in galaxies.
\bigskip

{\bf 1.11 D. Miralles Caballero}{\it (CAB CSIC, Spain)}

\vskip0.2cm

{\it Characterization of star forming regions in (U)LIRGs}

\vskip0.2cm

A significant fraction (U)LIRGs are known to constitute interacting and merging systems, where star formation is triggered within the galaxies and along the tidal features that usually form. A systematic analysis of almost 3000 star forming regions in a representative sample of 32 (U)LIRGs has been performed by means of high angular resolution ACS/HST B and I images. This talk presents the results of the photometric characterization of these star forming regions as a function of the luminosity of the systems, the interaction phase and the distance to the closest nucleus. Characteristics such as sizes, colors and luminosities will also be compared with those of clusters observed in less luminous interacting galaxies such as the Antennae. 
\bigskip

 \subsection{{\bf Session 2: Triggering mechanisms and environments}}

{\bf 2.1 P. Di Matteo}{\it (Observatoire de Paris, France)}

\vskip0.2cm

{\it Starburst galaxies: triggering mechanisms and dependence on the environment}

\vskip0.2cm

Strong starbursts in the local Universe are generally concentrated into the central regions of galaxies. Internal dynamical processes, related to non axisymmetric perturbations like bars, spiral patterns and lopsidedness, play an important role in driving angular momentum redistribution  and determining gas inflows into the central regions. These non-axisymmetric perturbations in galaxy disks can be produced by secular evolution, through accretion from cosmological filaments, or during much more violent processes, like interactions and mergers. After reviewing the role played by asymmetries in driving gas inflows and star formation enhancements in the central regions of galaxies, I discuss the impact interactions and mergers have in enhancing star formation, and the resulting average intensity, frequency and duration of merger-driven starbursts. Finally, I present some recent works on the effects the environment has in triggering (or limiting) bursts of star formation, discussing in particular the evolution of galaxies in compact groups and in the periphery of galaxy clusters.
\bigskip

{\bf 2.2 L. Kewley}{\it (University of Hawaii, USA)}

\vskip0.2cm

{\it Extreme Starbursts in Merging Galaxies}

\vskip0.2cm

I present new results from our large multiwavelength survey of nearby infrared merging galaxies. We show that metallicity gradients in galaxy pairs provide a ``smoking gun" for large-scale gas inflows in merging galaxies and are intricately connected to extreme central bursts of star formation. At first close pass, gas inflows dramatically flatten metallicity gradients which do not recover until very late merger stages. We use optical, infrared and X-ray diagnostics to investigate the evolution of starburst and AGN activity as a function of merger progress. We show that the galaxies form a clear merger sequence where fuel both starburst and AGN are fueled, and where the AGN becomes increasingly dominant during the final merger stages of the most luminous IR objects. Our results indicate that identification of the ``diffuse merger" stage is critical for understanding the connection (if any) between starburst and AGN activity. In this stage, extreme starburst and AGN activity co-exist, yet the presence of compact radio cores indicative of AGN disappears. We discuss these results in terms of thermal free free absorption caused by large amounts of merger driven gas inflows.
\bigskip

{\bf 2.3 C. Ramos Almeida}{\it (University of Sheffield, UK)}

\vskip0.2cm

{\it The triggering mechanisms of powerful radio galaxies: mergers and interactions}

\vskip0.2cm

Despite speculation that both starburst and nuclear activity in galaxies may be intimately linked via the common triggering mechanism of mergers and interactions, very little is known about the true nature of the link. Thus, the role of AGN in the formation and evolution of galaxies is still not well established. I present deep Gemini/GMOS imaging observations which are used to investigate the triggering mechanism(s) in a complete sample of radio-loud AGN for which, uniquely, we have quantified the level of both the AGN and star formation activity. I show results on the proportion of powerful radio galaxies triggered in galaxy mergers and also on the link between the degree of star formation/AGN activity and the interaction status of the host galaxies.
\bigskip

{\bf 2.4 B. Weiner}{\it (Steward Observatory, USA)}

\vskip0.2cm

{\it Spatial clustering of infrared-luminous galaxies, a clue to their fates}

\vskip0.2cm

The most massive starbursts in both the local and high redshift universe manifest themselves as ultraluminous infrared galaxies. However, it remains controversial what IR-luminous galaxies at z=1 are, and what they will evolve into. Are IR-luminous galaxies at high redshift mostly galaxy mergers, as they are at low redshift? Are ultraluminous IR galaxies strongly clustered, and can we infer whether they must evolve into cluster galaxies today? We measure the spatial clustering of LIRGs and ULIRGs at z=1 using Spitzer/MIPS sources cross-correlated with galaxies from the DEEP2 redshift survey. Because the evolution of clustering strength is well understood, the correlation lengths constrain the galaxy populations that LIRGs and ULIRGs will evolve into, and the classes of AGN to which they may be linked.
\bigskip

{\bf 2.5 B. Rothberg}{\it  (Naval Research Laboratory, USA)}

\vskip0.2cm

{\it The Impact of Star-Formation and Gas Dissipation on Galaxy Kinematics}

\vskip0.2cm

Mergers in the local universe present a unique opportunity for studying the metamorphoses of galaxies in detail. Yet, many studies and simulations show gas-rich mergers do not contribute significantly to the overall star-formation rate and total mass function of galaxies. The ultimate implication is that Lambda-CDM and our current understanding of galaxy formation and evolution may be completely wrong. I discuss recent results, based on high-resolution imaging and multi-wavelength spectroscopy, which demonstrate how star-formation and the presence of multiple stellar populations has lead to a serious underestimation of the dynamical masses of star-forming galaxies, in particular, Luminous \& Ultraluminous Infrared Galaxies. The presence of Red Supergiants and Asymptotic Giant Branch stars can severely affect the global properties measured in a galaxy, including: mass, age, extinction, and star-formation rate. I also discuss the impact of these stellar populations on studies of high redshift galaxies.
\bigskip

{\bf 2.6 J.C. Tan}{\it (University of Florida, USA)}

\vskip0.2cm

{\it Star Formation: From Local Regions to Extreme Starbursts}

\vskip0.2cm

I review the star formation process as we understand it from studies of nearby Galactic regions, including infrared dark clouds, embedded clusters, and the local volume of the Milky Way. I then discuss how the various physical processes are expected to change as we go to more extreme proto-star clusters and more extreme circumnuclear disks.
\bigskip

{\bf 2.7 S. Jogee}{\it (University of Texas at Austin, USA)}

\vskip0.2cm

{\it Assembly Modes and Star Formation of Galaxies out to z$\sim$3}

\vskip0.2cm

Mergers, smooth accretion, and secular processes are relevant for the assembly and central activity of galaxies in hierarchical models of galaxy evolution, but their relative importance at different epochs remains hotly debated. I discuss the role of galaxy mergers on star formation and structural assembly based on three of our studies, which target galaxies from the local Universe out to redshifts of 3: (1) In Jogee et al. \& the GEMS collaboration (2009), we explore the frequency of galaxy mergers and their impact on star formation over the last 7 Gyr using HST ACS, COMBO-17, and Spitzer data from the GEMS survey. We also compare the empirical merger history for high mass galaxies to theoretical predictions from five different suites $\Lambda$CDM-based models. Among high and intermediate mass systems, we find that the mean SFR of visibly merging systems is only modestly enhanced compared to non-interacting galaxies, and that visibly merging systems only account for less than 30$\%$ of the cosmic SFR density over this interval. (2) In Weinzirl, Jogee, Khochfar, Burkert \& Kormendy (2009), we set constraints on the merger history of high mass systems out to z$\sim$2 based on the structural property of local bulges. (3) In Weinzirl, Jogee, and the GINS collaboration (2010, in prep.), we discuss the structure, very high star formation rate, and AGN activity of the most massive galaxies (M$\star=5\times10^{10}-{\rm few}\times10^{12}$ M$_{\odot}$) at redshifts of z$\sim$2-3, and discuss the implications for galaxy evolution models.

\bigskip

\newpage

 \subsection{{\bf Session 3: Star formation histories}}

{\bf 3.1 R. Gonz\'alez Delgado}{\it (IAA CSIC, Spain)}

\vskip0.2cm

{\it Stellar Populations models: testing stellar libraries for Starbursts}

\vskip0.2cm

We revise the relevance of the stellar libraries for fitting the optical stellar continuum of a young and intermediate age components which are the dominant stellar populations in starbursts at optical wavelengths. The most up-to-date stellar libraries (Granada library, MILES, STELIB) are used in the evolutionary synthetic models for the spectral synthesis of stellar clusters in the LMC/SMC. These clusters were chosen because, due to their ages and metallicities, they have prominent Balmer absorptions, like many local starbursts. The comparison of the results obtained for the stellar clusters derived from this method and previous estimations available in the literature based mainly in the CMD method allow us to evaluate the pros and cons of each set of libraries to determine the age, metallicity and extinction for a stellar population. These comparisons allow us to estimate the impact of the uncertainties of the analysis of optical continuum spectra to estimate the ages and metallicities of starbursts.
\bigskip

{\bf 3.2 J. Rodr\'\i guez-Zaur\'\i n}{\it (CAB CSIC, Spain)}

\vskip0.2cm

{\it Star formation histories and evolution in local LIRGs and ULIRGs}

\vskip0.2cm

Luminous and Ultraluminous infrared galaxies (LIRGs and ULIRGs) are much more numerous at higher redshifts than locally, dominating the star formation rate density at redshifts $\sim$2. Therefore, they are important objects in order to understand how galaxies form and evolve through cosmic time. Local samples provide a unique opportunity to study these objects in detail. With that in mind, we have undertaken a program of spectroscopic observations of over 40 objects involving both, long-slit and integral field spectroscopic datasets. Our aim is to investigate the distributions of the parameters associated with the stellar populations (i.e. age, reddening and percentage contribution), and also, whether the properties of the stellar populations correlate with other properties of (U)LIRGs (such as the spectral classification, infrared luminosity, etc...).  We find that the star formation histories of ULIRGs are complex, with at least two epochs of star formation activity and that the characteristic timescale of the star formation activity is $<$100 Myr. These results are consistent with models that predict an epoch of enhanced star formation coinciding with the first pass of the merging nuclei, along with a further, more intense, episode of star formation occurring as the nuclei finally merge together. It is also found that the young stellar populations (YSPs) tend to be younger and more reddened in the nuclear regions of the galaxies. This is in good agreement with the merger simulations, which predict that the bulk of the star formation activity in the final stages of mergers will occur in the nuclear regions of the merging galaxies. In addition, our results show that ULIRGs have total stellar masses that are similar to, or smaller than, the break of the galaxy mass function (m$\star = 1.4\times10^{11}$ M$_{\odot}$). The mass estimates increase by a factor or $\sim$2 when accounting for old ($\sim$10 Gyr) stellar populations. Finally, we find no significant differences between the ages of the YSP in ULIRGs with and without optically detected Seyfert nuclei, nor between those with warm and cool mid- to far-IR colors. While these results do not entirely rule out the idea that cool ULIRGs with HII/LINER spectra evolve into warm ULIRGs with Seyfert-like spectra, it is clear that the AGN activity in local Seyfert-like ULIRGs has not been triggered a substantial period ($\geq$100 Myr) after the major merger-induced starbursts in the nuclear regions.

\bigskip

{\bf 3.3 R. Cid Fernandes}{\it (Universidade Federal de Santa Catarina, Brasil)}

\vskip0.2cm

{\it Fossil methods applied to U/LIRGS: Challenges, Strategies and Results}

\vskip0.2cm

Spectral synthesis (the ``art" of retrieving the star formation history of galaxies by means of spectral analysis using the most up to date spectral models for stellar populations) progressed enormously in the past half decade or so, fostering equally large advances in our understanding of galaxy evolution. Systems with intense star formation and large amounts of dust, however, are still challengingly complex to model. This contribution illustrates these difficulties by applying modern spectral synthesis techniques to Luminous and Ultra Luminous Infra-Red galaxies. A new version of the code STARLIGHT was developed to handle these systems, which incorporates optical spectra plus Far-IR data in order to constrain the star formation history of these (mostly interacting) galaxies. Strategies to overcome the difficulties and degeneracies involved are presented. 
\bigskip

{\bf 3.4 M. Rodrigues}{\it (GEPI-Observatoire de Paris, France)}

\vskip0.2cm

{\it How to retrieve stellar populations in starbursts}

\vskip0.2cm

In starburst galaxies, the light emitted by massive stars dominates the photon budget along most of the spectral energy distribution: hidden by the luminous stars, the fraction of old stellar population is systematically underestimated by current methods (Wuyt et al. 2009). This systematic has a large impact on the study of stellar populations and stellar masses in distant galaxies, when galaxies were actively forming stars. We have recently implemented a new algorithm to retrieve stellar populations from spectroscopic and photometric data using constraints from the observed SFR. The method relies on a meta-heuristic minimization method (the swarm intelligence algorithm). It allows us to alleviate the well known degeneracy between age and extinction, and better extract the hidden older stellar populations.
\bigskip

{\bf 3.5 R. Delgado-Serrano}{\it (Observatoire de Paris, France)}

\vskip0.2cm

{\it Starburst galaxies in the present and past Hubble sequence}

\vskip0.2cm

The way galaxies assemble their mass to form the well-defined Hubble sequence is amongst the most debated topic in modern cosmology. One difficulty is to link distant galaxies to those at present epoch. One observational way is at establishing how were the galaxies of the past Hubble sequence (e.g., 6 Gyr ago). We have derived a past Hubble sequence that can be causally linked to the present-day one. We selected a sample of nearby galaxies from the SDSS and a sample of distant galaxies from the GOODS survey. We verified that each sample is representative of the respective galaxy population. We further showed that the observational conditions necessary to retrieve their morphological classification are similar in an unbiased way. Galaxies are also divided into starburst and quiescent, according to whether EW([OII]$\lambda$3727) is larger or smaller than 15 \AA, respectively. After morphological classification we have been able to quantify their number fractions. We found an absence of Lumber evolution for elliptical and lenticular galaxies, which strikingly contrasts with the strong evolution of spiral and peculiar galaxies. Spiral galaxies were 2.3 times less abundant in the past, that is exactly compensated by the strong decrease by a factor 5 of peculiar galaxies. It shows that more than half of the present-day spirals had peculiar morphologies 6 Gyr ago, and this has to be accounted for by any scenario of galactic disk evolution and formation. The past Hubble sequence can be used to test these scenarios as well as to test evolution of fundamental planes for spirals and bulges.

\bigskip

\subsection{{\bf Session 4: Feedback effects}}

{\bf 4.1 C. Tadhunter}{\it (University of Sheffield, UK) }

\vskip0.2cm

{\it AGN feedback: when, how, and how much?}

\vskip0.2cm

Despite that general importance for understanding the evolution of massive galaxies, we still understand relatively little about AGN-driven outflows and the impact they have on their host galaxies. Concentrating on samples of radio-loud AGN, I review the following aspects: the triggering and timing of AGN activity; the link between AGN and starbursts; the observational evidence for AGN outflows; and the energetic significance of AGN-induced outflows compared with those driven by starbursts.

\bigskip

{\bf 4.2 H. Spoon}{\it (Cornell University, USA)}

\vskip0.2cm

{\it Mid-infrared kinematic evidence for outflows in ULIRGs}

\vskip0.2cm

Abundant evidence exists for the presence of outflows in active and starburst galaxies. Ultraluminous infrared galaxies host both extreme starbursts and supermassive black holes accreting at high Eddington rates. Outflows are hence expected to be ubiquitous in ULIRGs. The mid-infrared wavelength range is home to a suite of strong fine-structure lines from various ionization stages of neon, spanning a range in ionization potentials of 21 to 127eV, relatively unaffected by extinction. These lines can be exploited to probe the ionization structure and origin of the outflowing gas. We report on the results of a first systematic study of the line profiles of the mid-infrared fine-structure lines of Ne$^+$, Ne$^{2+}$, Ne$^{4+}$ and Ne$^{5+}$ gas in a sample of 320 ULIRGs, HyLIRGs, Seyferts, LINERs, QSOs and starburst galaxies observed by Spitzer-IRS. The sources span a range of 5 decades in [NeV] AGN luminosity and 6 decades in 21cm radio luminosity. Blue shifted [Ne III] and/or [Ne V] emission (shifted by 200 km/s or more) is found for 30$\%$ of the ULIRG sample. The incidence of blue shifted [Ne V] emission is even higher (60$\%$) among the sources with a [Ne V] detection. A comparison of the line profiles of the neon lines reveals the ionization of the blue shifted gas to increase with blue shift, implying decelerating outflows in a stratified medium, photo-ionized by the AGN.
\bigskip

{\bf 4.3 P. Hopkins}{\it (University of California Berkeley, USA)}

\vskip0.2cm

{\it Gas, Galaxy Mergers, Starbursts, and AGN: Powering an Evolving Hubble Sequence}

\vskip0.2cm

In the last few years, the combination of models that include realistic large gas supplies in galaxies, and prescriptions for feedback from both stellar evolution and super-massive BHs to maintain those gas reservoirs, have led to huge shifts in our understanding of galaxy formation. In particular, gas-richness, and the magnitude of starbursts driven by tidal action, may represent the most important driving factor in the net effects of galaxy-galaxy mergers on bulge structural properties, stellar populations, mass profiles, and kinematics; models with the appropriate gas content have finally begun to produce realistic bulges that resolve a number of discrepancies with observations. In the regime of very gas-rich mergers, expected at high redshift and/or low masses, gas can qualitatively change the character of mergers and starburst galaxies, making disks robust to destruction in mergers and providing a natural explanation for the observed morphology-mass relation. These processes provide a link between the ``relic" population seen today, low-redshift starburst populations, and rapidly star-forming galaxies at high redshifts. Feedback is critical in a number of ways: it regulates and maintains gas supplies, can ``shut down" the tail-end of starburst activity leaving ``quenched" galaxies, and may set a characteristic upper limit to the densities reached by any rapidly star-forming systems from the scales of star clusters to the most massive high-redshift starbursts.
\bigskip

{\bf 4.4 D. Rupke}{\it  (Institute for Astronomy, University of Hawaii, USA)}

\vskip0.2cm

{\it Gas flows in Starburst Mergers}

\vskip0.2cm

Starbursts in merging galaxies are bookended by gas flows. They are preceded by strong radial inflows that fuel the star formation, and followed by outflows of enriched and entrained gas that act as regulatory feedback. Using both observations and simulations, I present new results on (1) how we constrain gas inflows by tracking metals in starburst mergers and (2) how we are finally revealing the complex structures of the ubiquitous gas outflows in merging galaxies.  

\bigskip

\subsection{{\bf Session: Posters}}

{\bf P.1 A. Alonso Herrero}{\it (CAB CSIC, Spain)}

\vskip0.2cm

{\it PMAS Optical Integral Field Spectroscopy of Luminous Infrared Galaxies}

\vskip0.2cm

The general properties (activity class, star formation rates, metallicities, extinctions, average ages) of local luminous infrared galaxies (LIRGs) are well known since large samples have been the subject of numerous spectroscopic works. There are, however, relatively few studies of large samples of
LIRGs using integral field spectroscopy (IFS). We present optical (3800-7200 \AA) IFS data taken with the Potsdam Multi-Aperture Spectrophotometer (PMAS) of the central few kiloparsecs of eleven LIRGs. We complemented the PMAS observations with existing HST/NICMOS Pa$\alpha$ imaging.   The optical continua of selected regions are well fitted with combinations of evolved (0.7-10 Gyr) and ionizing (1-20 Myr) stellar populations. The latter is more obscured than the evolved population, and has visual extinctions in good agreement with those obtained from the Balmer decrement. Except for NGC 7771, we found no evidence for an important contribution to the optical light from an intermediate-aged stellar population (100-500 Myr). Even after correcting for stellar absorption, a large fraction of spaxels with low equivalent widths of H$\alpha$ in emission still show enhanced [NII]6584/H$\alpha$ and  [SII]6717,6731/H$\alpha$ ratios. These ratios are likely to
be produced by photoionization in HII regions and diffuse emission. These regions of enhanced line ratios are coincident with low surface brightness HII regions. The fraction of diffuse emission in LIRGs varies from galaxy to galaxy, and it is less than
60$\%$ as found in other starburst galaxies.  The H$\alpha$ velocity fields over the central few kpc are generally consistent, at least to first order, with rotational motions. The velocity fields of most LIRGs are similar to those of disk galaxies, in contrast to the highly perturbed fields of most local, strongly interacting ULIRGs. The peak of the H$\alpha$ velocity dispersion coincides with the position of the nucleus and is likely to be tracing mass.
\bigskip

{\bf P.2 E. Bellocchi}{\it (CAB CSIC, Spain)}

\vskip0.2cm

{\it Kinemetry of local (Ultra)luminous Infrared Galaxies}

\vskip0.2cm

The kinematics of a sample of 42 (Ultra)Luminous Infrared Galaxies [(U)LIRGs] at low redshift (z$\leq$0.022) will be analyzed thanks to Integral Field Spectroscopy (IFS) carried out with the VIMOS instrument on the VLT. Studying the characteristics of (U)LIRGs at low redshift allow us a better understanding the interrelated physical processes involved, and the implications for high-z since these galaxy populations are more numerous at cosmological distances than locally. As preliminary steps, the data reduction and post-reduction have been performed using the EsoRex pipeline (by ESO) and IDL and IRAF scripts (creating the final data cube). Then, the line profiles of these spectra (e.g., H$\alpha$) will be studied in order to extract the relevant emission line information (central wavelength, FWHM and flux intensity). We are going to apply the kinematic criteria used by Shapiro et al. (2008) to the whole sample in order to distinguish galaxies dominated by ordered rotational motions (i.e., disk) and those involved in major merger events (i.e., merger) considering the asymmetries in both the velocity field and velocity dispersion maps of the warm gas (as traced by H$\alpha$). To this aim the ``kinemetry" method, developed by Krajnovi\'c et al. (2006), is considered. Here, the results of a sub-sample is discussed.
\bigskip

{\bf P.3 C. Cortijo Ferrero}{\it (IAA CSIC, Spain)}

\vskip0.2cm

{\it The LIRG galaxy IC1623: the nearest Lyman Break Analog}

\vskip0.2cm

We present the analysis of imaging data from Hubble Space Telescope (STIS, ACS and NICMOS) and Spitzer Space Telescope (IRAC bands), and the long-slit spectra from ALFOSC-NOT for the z=0.02 merging system IC1623. IC1623 is a very Luminous Infrared system, log(L$_{IR}$/L$_{\odot}$)=11.65, being the west component very bright at ultraviolet wavelengths too. The images show that the merging system is composed of two galaxies. The western component is very bright in the UV and optical, while the eastern one is very bright in the infrared. The west galaxy shows properties very similar to the cosmological Lyman Break galaxies, and it can be considered as the nearest Lyman Break Analog. We have identified the stellar clusters in the HST images, and the colours indicate that there are two different types of clusters: those located in the UV bright galaxy which have ages between 1-10 Myr and are little affected by extinction, less than 1 mag, and those located in the IR bright galaxy much older and extinguished. The masses of the clusters span a range $10^6-10^8$ M$_{\odot}$. The long-slit optical spectra through IC1623 are also analyzed with the code {\it Starlight} to estimate the properties (age, extinction and metallicity) of the nucleus, some of the brightest clusters and the underlying galaxy.
\bigskip

{\bf P.4 J. Piqueras L\'opez}{\it (CAB CSIC, Spain)}

\vskip0.2cm

{\it A  2D near-infrared study of a sample of local LIRGs and ULIRGs}

\vskip0.2cm

We are carrying out a large program involving integral field spectroscopy observations of local LIRGs and ULIRGs, with the aim of studying their kinematics, ionization mechanisms, etc. In particular, this study is based on VLT-SINFONI observations covering both H (1.45-1.85 $\mu$m) and K (1.90 - 2.50 $\mu$m) bands, with an intermediate spectral resolution (R$\sim$3000-4000), a FOV of $\sim$8"$\times$8", and spatial resolution of $\sim$0.25 arcsec/pixel. Here I present SINFONI reconstructed maps of the Hydrogen recombination lines (Pa$\alpha$ and the Brackett series), molecular Hydrogen excitation lines and metal lines (HeI, [FeII], [SiVI], etc) of a sample of 9 LIRGs and 7 ULIRGs. Since the reddening is less important at infrared wavelengths, we can investigate in detail the different ionizing mechanisms and how they relate with the star formation activity. A further study of the kinematics and the spatial distribution of the dust will prove the full potential of our dataset.

\bigskip

 \subsection{{\bf Session 5: The AGN/Starburst connection}}

{\bf 5.1 G. Canalizo}{\it (University of California Riverside, USA)}

\vskip0.2cm

{\it The Connection Between Starbursts and QSO Activity}

\vskip0.2cm

Recent observations of low redshift quasar host galaxies indicate that mergers and significant episodes of star formation are ubiquitous in these galaxies.  By constraining the timescales of such events we can gain a better understanding of the role of AGN feedback in galaxy evolution.  I will discuss results from a long campaign of space- and ground-based imaging and spectroscopic observations of z $<$ 0.5 hosts that imply that mergers are indeed essential for the triggering of quasars, and that these mergers invariably induce starbursts either during or shortly after the merger.   There appears to be, however, a large range of values for the time delays between the merger and the onset of the nuclear activity, varying from a few Myr to more than a Gyr.  We find some evidence for a bimodal distribution, although this could be a selection effect.
\bigskip

{\bf 5.2 M. Imanishi}{\it (National Astronomical Observatory, Japan)}

\vskip0.2cm

{\it The AGN-starburst connections in nearby luminous infrared galaxies}

\vskip0.2cm

We present the results of our systematic search for optically elusive, but intrinsically luminous buried AGNs in $>$100 nearby (z $<$ 0.3) luminous infrared galaxies with L(IR) $>$ 10$^{11}$ L$_{\odot}$, classified optically as non-Seyferts. To disentangle AGNs and stars, we have performed 
(1) infrared 2.5-35 $\mu$m low-resolution (R $\sim$ 100) spectroscopy using Subaru, AKARI, and Spitzer, to estimate the strengths of PAH (polycyclic aromatic hydrocarbon) emission and dust absorption features, 
(2) high-spatial-resolution infrared 20 $\mu$m imaging observations using Subaru and Gemini, to constrain the emission surface brightnesses of energy sources, and 
(3) millimeter interferometric measurements of molecular gas flux ratios, which reflect the physical and chemical effects from AGNs and stars. Overall, all methods provided consistent pictures. We found that the energetic importance of buried AGNs is relatively higher in galaxies with higher infrared luminosities (where more stars will be formed), suggesting that AGN-starburst connections are luminosity dependent. Our results might be related to the AGN feedback scenario as the possible origin of the galaxy down-sizing phenomenon. 
\bigskip

{\bf 5.3 M. Lazarova}{\it (University of California Riverside, USA)}

\vskip0.2cm

{\it Infrared SEDs and Star Formation Rates of LoBAL QSOs}

\vskip0.2cm

Low-ionization Broad Absorption Line QSOs (LoBALs) are a rare class of objects, accounting only for 1-3$\%$ of the general population of QSOs. Their defining characteristic is the presence of high velocity ($>$2000 km/s) mass outflows of low- and high-ionization ions, which are evident in the very broad blue-shifted absorption troughs in their rest-UV spectra. There is some observational evidence that LoBALs at low redshifts might exclusively reside in Ultra Luminous Infrared Galaxies (ULIRGs) with disturbed morphologies and young stellar populations as a result of a recent galaxy merger. Those studies and the currently sparked interest in AGN feedback as a possible mechanism for regulating galaxy evolution have highlighted the importance of testing previous ideas proposing that BALs represent a short-lived outflow phase early in the life of QSOs. Herein we present the first results from a multiwavelength, systematic study of a complete sample of 22 LoBALs drawn from the SDSS DR3 within 0.5 $<$ z $<$ 0.6. 
We model their optical through far-infrared SEDs using SDSS photometry, Spitzer/IRS low-resolution spectra from 7-20 $\mu$m and Spitzer/MIPS observations at 24, 70, and 160 $\mu$m. We estimate the total IR luminosities, star formation rates, and relative AGN/Starburst contribution to the FIR emission. We find that only half of the LoBALs in our sample reside in ULIRGs or HyLIRGs, while the rest of the hosts are LIRGs. We also estimate that the AGN accounts for 20-70\% of the FIR luminosity. Using only
the starburst contribution to the FIR luminosity, we estimate SFRs $\sim$40-300 solar masses per year, values typical of LIRGs. In order to interpret our results, we need a control sample of classical type 1 QSOs analyzed in the same way.

\bigskip

{\bf 5.4 M. Villar Mart\'\i n}{\it (IAA CSIC,  Spain)}
 
\vskip0.2cm

{\it Extended emission line nebulae around type 2} 

\vskip0.2cm

I present recent results based on VLT and GTC spectroscopic and imaging data of a new program whose main goal is to investigate the existence of extended emission line nebulae associated with type 2 quasars, characterize their properties and use them as cosmological tools.
\bigskip

{\bf 5.5 M. Brotherton}{\it (University of Wyoming, USA)}

\vskip0.2cm

{\it Post-Starburst Quasars}

\vskip0.2cm

I discuss our investigations to find and understand post-starburst quasars, hybrid objects with accreting black holes and recent massive starbursts, and their role in the evolution of galaxies. Data from SDSS, Keck, HST, Galex, and Spitzer are featured.
\bigskip

{\bf 5.6 V. Wild}{\it (IAP, France)}

\vskip0.2cm

{\it Timing the starburst AGN-connection}

\vskip0.2cm

There are many theories successful in explaining the observed correlations
between black holes and their host galaxies. In turn, these theories play a crucial
role in explaining other observed aspects of the galaxy population, such as the red/blue bimodality. However, observational measurements of the interaction of black
holes with their hosts remain scarce. I present the time-averaged growth of black
holes in the 400 strongest starbursts in local galactic bulges observed in the SDSS.
These bulges have experienced a strong burst of star formation in the past 600 Myr,
as indicated by the strength of the Balmer absorption lines in their integrated
stellar spectra. We select our sample of starbursts by fitting stellar population
models to spectral indices which measure the strength 4000 \AA\  break, shape of the
blue continuum, and Balmer absorption lines. The sample is complete in the sense
that there is an equal number of starburst galaxies per unit age, up to an age of
600 Myr after the starburst.
We use the luminosity of [OIII]$\lambda$5007 \AA\ to measure the black hole accretion rate in
galaxies identified as having an AGN via the BPT diagram, carefully correcting for
dust attenuation and contamination from star formation. Averaging over the whole
sample of 400 galaxies, we find that there is a delay of 250 Myr between the onset
of the starburst and accretion onto the black hole. This coincides with the time
when the mass-loss rate from supernova and OB star winds has declined to a level
below the mass-loss rate from less massive stars. However 250 Myr is significantly
later than the onset of mass-loss from low-mass stars (about 50 Myr). We suggest
that mass ejected in fast supernova and OB winds is not accreted onto the black
hole, and further that supernovae may prevent the accretion of mass onto the black
hole between 50-250 Myr after the starburst.
The luminosity of H$\alpha$ provides an entirely independent measure of the
instantaneous star formation in the sample, and a detailed picture of the evolution
of star formation rate after a starburst. We find that the SFR declines slowly, over a
typical timescale of 300 Myr. We also observe 3 distinct phases of evolution of the
star formation rate: (1) the ``starburst'' phase, a sharp spike lasting a few tens of
Myr; (2) a ``coasting" phase, the star formation rate remains roughly constant up to
400 Myr; (3) the ``post-starburst'' phase, the star formation decays rapidly.
With both bulge and black hole increasing their mass by $\sim$10\% in 600 Myr, the
processes at work in this local starburst sample may well be relevant to the coevolution
of black holes and bulges over cosmic time. Extrapolating a simple model
in which the black hole accretes $\sim$0.5\% of the mass ejected from low-mass stars
formed during the starburst after 250 Myr, leads to a bulge:black hole mass ratio of
400, remarkably close to the value of 700 observed in local bulges today.
\bigskip

\subsection{{\bf Session 6: The distant universe}} 

{\bf 6.1 M. Swinbank}{\it (Durham University, UK)}

\vskip0.2cm

{\it The Properties of star-forming regions in high-z star-forming galaxies}

\vskip0.2cm

Measuring the properties of star-forming regions in high redshift galaxies (such as sizes, luminosities, and velocity dispersions) define some of the key science drivers for ELT and ALMA. Such observations can tell us how and why the star-formation in distant galaxies is much for efficient than that seen locally, and whether local, intense star-forming regions are good analogs for high-z galaxies. In this talk, I will show some recent observations which have been aided by strong gravitational to probe the properties of star-forming regions within galaxies at z$\sim$2--5 on scales of $\sim$100pc. These results show that the mode of star-formation at z$\sim$2 is similar to that seen in local ULIRGs, although the energetics are unlike anything seen in the local Universe.
\bigskip

{\bf 6.2 K. Menendez-Delmestre}{\it (Carnegie Observatoires, USA)}

\vskip0.2cm

{\it An IFU view to Extreme Starbursts at z$\sim$2: the Case of Submm Galaxies}

\vskip0.2cm
Ultra-luminous infrared galaxies (ULIRGs) are locally rare, but appear to dominate the co-moving energy density at z$>$2. Many are optically-faint, dust-obscured galaxies that have been identified only recently by the detection of their thermal dust emission redshifted into the submillimeter wavelengths. These submm galaxies (SMGs) have been popular candidates to be progenitors of the most massive galaxies at z$\sim$0. With colossal ULIRG-like luminosities that translate into unusually high SFRs ($\sim$100-1000 M$_{\odot}$/yr), SMGs could build the stellar bulk of a massive galaxy in under a few hundred million years. However, the predominance of AGN signatures in these SMGs shows that star formation and AGN activity coexist in these objects, implying that we are witnessing the coupled growth of the stellar spheroid and a central SMBH.
We have undertaken the first integral-field spectroscopic observations aided with adaptive optics (AO) of SMGs. With the OSIRIS integral field unit (IFU), designed to be used with the Keck Laser Guide Star Adaptive Optics (LGS-AO) system, we investigate the distribution of H$\alpha$ line emission in 3 SMGs at 1.4$<$z$<$2.4. LGS-AO allows us to probe down to kpc-scale spatial resolutions, up to 10 times more resolved than what prior seeing-limited observations had been able to achieve. The exquisite resolution provided by LGS AO allows us to spatially distinguish between AGN and star-forming regions as revealed by differences in H$\alpha$ spectral properties and to uncover velocity offsets ($\sim$few$\times$100 km/s) between individual galactic-scale sub-components. We find that, after an estimated correction for extinction based on typical Balmer decrements for SMGs, their high SFR surface densities are similar to local extremes like ULIRGs and starbursts. However, their spatial extensions stretch beyond $>$8 kpc, suggesting that SMGs may be undergoing such intense star-forming activity on significantly larger spatial scales than extreme local environments, which are typically concentrated in $\sim$1-2 kpc.
\bigskip

{\bf 6.3 F. Hammer}{\it (GEPI-Observatoire de Paris, France)}

\vskip0.2cm

{\it Disk formation in massive spirals: merger or secularly induced star formation?}

\vskip0.2cm

Using the deepest and most complete observations of distant galaxies, we investigate how large disks could have been formed. Observations include spatially-resolved kinematics, detailed morphologies and photometry from UV to mid-IR. Six billions years ago, half of the present-day spirals were experiencing major mergers, evidenced by their anomalous kinematics and morphologies as well as their relatively high gas fractions. They are consequently modeled using the state of the art hydrodynamics models. This provides a new channel of disk formation, e.g. disks reformed after gas-rich mergers. Then one may estimate which fraction of the stellar mass density has been formed during mergers. This will be compared to expectations from nearby galaxies, including the Milky Way and M31.
\bigskip

{\bf 6.4 P. G. P\'erez-Gonz\'alez}{\it (UCM, Spain)}

\vskip0.2cm

{\it Understanding the mass assembly of galaxies at 0$<$z$<$4: Spitzer's contribution and open questions}

\vskip0.2cm

We present the main results of our research about the assembly of galaxies at z$<$4 based on the data obtained by the deepest Spitzer surveys carried out with IRAC and MIPS during the cryogenic mission. These data in the near-, mid- and far-IR have allowed us to obtain unprecedentedly robust estimations of the obscured SFRs and stellar masses of distant galaxies. Analyzing SFR and stellar mass functions in several redshift bins at 0$<$z$<$4, we have found and quantified that galaxies formed following a downsizing scenario, with the most massive systems assembling early in the lifetime of the Universe and very quick (i.e., with very high star formation efficiencies, and a significant amount of obscured starbursts), while less massive systems assembled later and/or more slowly. However, Spitzer has left several open questions that still hamper our current understanding about the formation and evolution of galaxies. I discuss three of these results and how future facilities such as Herschel, ALMA, E-ELT or JWST can lead to a more robust and detailed (with higher spatial resolution and depth) characterization of how galaxies formed in the early Universe: (1) the mid-to-far IR colors of galaxies evolve with redshift, departing considerably from the typical values observed in the local Universe, specially at z$>$1.5-2.0; (2) the IMF might not be universal, evolving to a top-heavy IMF at z$>$1.5; (3) obscured AGN may be ubiquitous in high-z galaxies, playing a significant role in the downsizing scenario.
 
\bigskip

\subsection{{\bf Session 7: Summary and future prospects}}

{\bf 7.1 J. Turner}{\it (UCLA, USA)}

\vskip0.2cm

{\it Molecular Gas and Star Formation in Starbursts: A Closer Look}

\vskip0.2cm

Molecular clouds are the fuel for starbursts. What conditions cause molecular clouds to create starbursts? Are high gas surface densities sufficient? What effects do starbursts have on surrounding molecular gas? The current state of our knowledge of star formation and gas links are reviewed, and extrapolated to what upcoming observations in the far-IR and millimeter/submillimeter may reveal about the causes and effects of starbursts.
\bigskip

{\bf 7.2 J. Fischer}{\it (Naval Research Laboratory, USA)}

\vskip0.2cm

{\it Herschel PACS Spectroscopy of ULIRGs}

\vskip0.2cm

I describe our Herschel PACS survey of local Ultraluminous Infrared Galaxies (ULIRGs), which is part of the SHINING Guaranteed Time survey of local galaxies. In particular, I discuss far-infrared spectroscopy of Mrk 231, the most luminous of the local ULIRGs, and a type 1 broad absorption line AGN. For the first time in a ULIRG, all observed far-infrared fine-structure lines in the PACS range were detected and all were found to be deficient relative to the far infrared luminosity by one to two orders of magnitude compared with lower luminosity galaxies. The deficits are similar to those for the mid-infrared lines, with the most deficient lines showing high ionization potentials. Aged starbursts may account for part of the deficits, but partial covering of the highest excitation AGN powered regions may explain the remaining line deficiencies. A massive molecular outflow, discovered in OH and $^{18}$OH, showing outflow velocities out to at least 1400 km/s, is a unique signature of the clearing out of the molecular disk that formed by dissipative collapse during the merger. The outflow is characterized by extremely high ratios of $^{18}$O/$^{16}$O suggestive of interstellar medium processing by advanced starbursts.
\bigskip

{\bf 7.3 S. F. S\'anchez}{\it (CAHA)}

\vskip0.2cm

{\it CALIFA: Calar Alto Legacy IFs Astronomical Survey}

\vskip0.2cm

We present CALIFA, an IFS survey of $\sim$600 galaxies in the local Universe (z$<$0.03), to be performed with PPAK@3.5m telescope at Calar Alto, aimed to study the spatial resolved properties of the stellar populations and ionized gas within the $\sim$90$\%$ of the area covered by the galaxies, by sampling the optical wavelength range between 3700-7100 \AA\  with a resolution of R$\sim$1000/2000. The main goals of this survey would be to understand the details of the star formation history, galaxy growth and evolution within the Hubble sequence, fixing the anchor point of the cosmological evolution of galaxies.
\bigskip

{\bf 7.4 L. Colina}{\it (CAB CSIC, Spain)}

\vskip0.2cm

{\it Mid-IR JWST view of dusty starbursts near and far}

\vskip0.2cm

I present MIRI, the Mid-IR imager and spectrograph for the JWST. MIRI will offer unique capabilities for the study of the dustiest and more extreme starbursts in the local Universe and at cosmological distances.

\bigskip

\newpage

\section{Summary and Conclusions of the Workshop by S. Veilleux}

In this summary paper, the many interesting topics addressed or
discussed at this workshop are divided into four main themes: (1) the
life and death of local starbursts, (2) feedback, (3) mergers:
observations versus simulations, and (4) the more distant universe.
The key open issues under each theme are pointed out and the prospect
in the next 10 years to solve these issues is discussed, emphasizing
upcoming space missions and ground-based facilities that will greatly
help with this task.

\vskip0.2cm

\noindent
Overall, 41 talks and 4 posters, with their respective 10-minute short
talks, were presented over a period of 4.5 days. In addition, there
were 6 discussion sessions of up to one hour during the workshop.
A wide variety of topics were addressed or discussed; they are
summarized below along with some of the key open issues. Throughout
this summary, the names of the speakers who discussed the
corresponding topics are indicated in brackets; the readers should
refer to the original papers for more details. Text in italics
indicates the open issues and unanswered questions. 

\subsection{{1.The life and death of local starbursts}}

\noindent
One of the central topics of this workshop was the quantitative study of
local starbursts to derive their basic properties (e.g., star formation
rate, efficiency, history) in relation to the properties of their
host galaxies (e.g., stellar masses, kinematics). 

\subsubsection{{The tracers of the star formation activity (and AGN contamination)}}

\noindent
The first aspect of starburst studies is to trace and quantify star
formation activity. In principle, this can be done using several 
spectral features from the UV to the far-IR and radio. However, each
of the tracers has its advantages and drawbacks. Here are the tracers
of star formation activity that were discussed during the workshop and
the objects where these tracers were applied:

\begin{itemize}
\item Optical: BPT/VO87/Kewley's line ratio diagnostics diagrams, used
  to quantify starburst/AGN activity in ULIRGs. {\em Dust
    obscuration?} [L. Kewley]
\item Near-IR: PAH 3$\mu$m feature in ULIRGs. {\em Dust
    obscuration?}  [M. Imanishi]
\item Spitzer mid-IR (MIR): low/high-ionization fine structure lines, PAHs,
  24$\mu$m to estimate star formation rates (SFR), excitation
  temperatures (T$_{exc}$), dependence on metallicity and ionization
  index in a wide variety of galaxies [A. Alonso-Herrero]. MIR extents
  are smaller among (U)LIRGs with dominant AGN and warmer 60/100$\mu$m
  colors [T. D\'iaz Santos].
\item
  Far-IR: plethora of fine structure and molecular lines such as CO, HCN....\\
  {\em Concerns: IR pumping/X-rays affect the chemistry of the
    interstellar medium; the ratio of MIR and FIR lines-to-FIR
    continuum in ULIRGs $\sim$~(1 -- 10\%)$\times$ lower-luminosity
    galaxies (giant wiffle ball geometry? origin of FIR continuum in
    these objects?)} [S. Grac\'ia Burillo, J. Turner, J. Fischer].
\item Radio: Merlin + eEVN, to detect/monitor individual supernovae
  (SNe) with $\Theta \sim$ few mas, direct measurement of the supernova
  rate, hence SFR, less affected by obscuration ({\em possible
    confusion with UCHII regions?})  [A. Alberdi, M. P\'erez Torres]
\end{itemize}

\subsubsection{Star formation laws and (in)efficiency}

\noindent
One important question is whether the same basic physical processes
governing star formation activity on the small local scale also apply
to galactic scale and more extreme environments like that in
(U)LIRGs. The following topics were discussed within this context during
the workshop:

\begin{itemize}
\item
  Theoretical/numerical studies: \\\\
  -- GMC collisions/mergers compress gas that is already molecular
  $\rightarrow$ shear-driven GMC collision law to explain star
  formation activity from normal galaxies to circumnuclear starbursts
  [J.C. Tan]
\item
  Observational studies:\\\\
  -- The Milky Way and nearby galaxies represent a convenient ``sanity check", thanks to excellent spatial resolution they provide.  A broad range of star formation efficiency, SFE, is observed: $\sim$1-10\%.  {\em Difficult to form SSCs. } [J. Turner]\\\\
  -- (U)LIRGs: SFE(merger)$_{dense HCN} \sim$ 3 $\times$ SFE(normal)$_{dense HCN}$, similar to high-z CO observations (e.g., Genzel et al., 2010) [S. Garc\'ia Burillo]\\\\
  --{\em Major uncertainties include: X$_{CO}$, X$_{HCN}$, SFR, AGN
    contamination.}
\end{itemize}

\subsubsection{Stellar masses and star formation histories}

\noindent
The approach of fitting stellar population models to photometric
points and/or spectroscopic data is commonly used to study the star
formation histories and evolution of star-forming galaxies. The
advantages and caveats associated with the various techniques were
broadly discussed during the workshop along with examples of
applications on actual galaxies.

\begin{itemize}
\item To quantify the uncertainties on age and metallicity, stellar
  libraries/evolutionary synthetic models are applied to well-known
  stellar clusters in LMC and SMC [R.M. Gonz\'alez Delgado].
\item Additional constrains are included in the modeling to reduce
  the impact of degeneracies between the different parameters
  (e.g., age and reddening). (1) Take into account the FIR in (U)/LIRGs
  [the {\it Starlight} code, R. Cid Fernandes]. (2) Constrain models
  to give SFR$_{tot}$ = SFR$_{UV}$ + SFR$_{IR}$ $\rightarrow$
  alleviate age-extinction degeneracy and better extract older stellar
  populations [M.\ Rodrigues]
\item
  Applications:\\\\
  -- U/LIRGs: analysis suggests multiple epochs of star formation activity [R. Cid Fernandes]\\\\
  -- U/LIRGs: a useful simplification is to assume only young ($\leq$ 100 Myr;
  YSPs) + intermediate-age (0.1 $<$ age $\leq$ 2 Gyr; ISPs) populations
  because the old stellar populations do not contribute significantly
  to the optical continuum. YSPs are more important in the nuclear
  regions, where they tend to be younger and redder. No other obvious
  trends [J.\ Rodr\'iguez Zaur\'in]
\item {\em Caveats: uncertainties on metallicities, difficulties in modeling
    near-IR range, self-consistent treatment of continuum and emission
    lines from starbursts is lacking. }
\end{itemize}

\subsubsection{2D mapping of starbursts galaxies}

\noindent
Impressive emission-line imaging and integral-field spectroscopic
datasets were presented in various talks and posters. These data were used 
to investigate, for example, the kinematics, ionization mechanisms
and dust distribution in star-forming galaxies: 

\begin{itemize}
\item
Excitation/extinction/SSCs.\\\\
-- Calar alto/PMAS of ULIRGs: enhanced [N~II]/H$\alpha$, [S~II]/H$\alpha$ at low surface brightness 
[Poster by A. Alonso Herrero]\\\\
-- VLT/VIMOS of 32 LIRGs + 6 ULIRGs: enhanced [N~II]//H$\alpha$, [S~II]/H$\alpha$, [O~I]/H$\alpha$ correlated with interaction stage [A. Monreal Ibero]\\\\
-- LBT/LUCIFER: SFR, SNR/SFR, extinction maps of NGC1569 [A. Pasquali]\\\\
-- HST/ACS of SSCs in 32 U/LIRGs: $L$, $r_{eff}$ increase with $L_{\rm IR}$ and merger phase $\rightarrow$ select TDG candidates [D. Miralles Caballero].
\item
Kinematics (on-going):\\\\
-- VLT/VIMOS of 32 LIRGs and 6 ULIRGs: study of the kinematics using {\it kinemetry} (Shapiro et al., 2008) to distinguish disk- vs merger-dominated galaxies [Poster by E. Bellocchi]\\\\
-- VLT/SINFONI of 9 LIRGs and 7 ULIRGs: study of the kinematics of the systems as well as the different ionization mechanisms [Poster by J. Piqueras L\'opez].
\end{itemize}

\subsection{2. Feedback}

\noindent
Swept under the rug for many years, the dirty f-- word, feedback, is
now at the center of most discussions on galaxy evolution.  The role
of both AGN- and starburst-driven feedback on the surrounding ISM is a
key open issue. During the workshop there were lively discussions on
this topic and many interesting results were presented:

\begin{itemize}
\item
ULIRGs:\\\\
-- Neutral gas outflows in ULIRGs: Na~ID. Spatially extended. Highly energetic $\sim$ 10$^{56}$ -- 10$^{59}$ ergs. Most likely starburst driven. 2D IFU data reveal decoupling of neutral gas from ionized gas in one object [D. Rupke].\\\\
-- Ionized gas outflows in ULIRGs: MIR Ne lines. Blueshift increases with ionization level of blueshifted gas $\rightarrow$ decelerating outflows in stratified medium, photoionized by AGN. {\em Comparison between optical neutral wind and ionized wind?} [H. Spoon].
\item 
AGN-driven outflows:\\\\
-- Radio mode extends on galactic/ICM/IGM scales but is present only in $\sim$ 10\% of AGN. Quasar wind mode is present in many systems {\em but there is little evidence that it is energetically significant} [C. Tadhunter].\\\\
-- Powerful molecular outflow in Mrk 231, velocity matched with spatially resolved neutral wind detected by Rupke et al 2005 [J. Fischer]. 
\end{itemize}

\subsection{3. Mergers: observations vs simulations}

\noindent
Recent merger simulations often make an attempt to predict the
star formation histories and evolution of the merging systems, which
can in principle be compared against the observations. Therefore, one
of the aims of the workshop was to bring together theoretical and
observational astronomers to share ideas and better understand
the physics behind star formation activity in mergers and
interactions.

\begin{itemize}
\item Simulations: dependence of merger evolution on orbital
  parameters, mass ratios, gas fractions, feedback,
  environments,...[P. Di Matteo, P. Hopkins].
\item Gas inflows: reduce Z(center) and flatten dZ/dR [L. Kewley,
  D. Rupke]
\item Merger remnants: presence of red supergiants and AGB stars from
  starburst affect estimates of stellar mass, age, extinction,
  SFRs. {\em Dynamically decoupled cores?} (B. Rothberg). Excess light
  in Es $\rightarrow$ starburst fraction [P. Hopkins]
\item
  ULIRG evolution:\\\\
  (1) AGN becomes increasingly dominant along the merger sequence probed by ULIRGs. {\em What is the nature of composite systems/diffuse merger state?} [L. Kewley, M. Imanishi]\\\\
  (2) Spitzer/MIPS + DEEP2 redshift indicate that z$\sim$1 ULIRGs are
  strongly clustered. {\em possibly more than QSOs?} [B. Weiner].
\item Radio Galaxies (RGs): they show signs of interaction, but are at
  various merger stages. Often, starburst age $>$ RG age, implying a
  delay between RG and starburst [C. Ramos Almeida and C. Tadhunter]
\item IR-excess/LowBAL/Post-starbursts/Normal QSO: ditto! {\em Bimodal
    distribution of starburst ages? or selection effects?}
  [G. Canalizo, M. Lazarova, M. Brotherton]. Tidal features in some
  type 2 QSOs [M. Villar Mart\'in]. Delay of $\sim$250 Myr among SDSS
  bulge galaxies with $<$600 Myr starbursts [V. Wild].
\end{itemize}

\subsection{4. A long time ago in a galaxy far, far away...}

\noindent
A tremendous amount of work is being done on star-forming galaxies at
high redshifts, where it is necessarily more difficult to study in detail the
physical processes associated with star formation.
Therefore, another goal of the workshop was to bring together the
``high-z'' and ``local'' astronomical communities to share knowledge and
better understand the links between the local and high-z star-forming
galaxies.

\subsubsection{Local starbursts as templates of distant starbursts}

\noindent
A number of talks and one of the discussions addressed the
possibility of using local analogues to study distant starbursts:

\begin{itemize}
\item
  Local analogues to Lyman Break Galaxies (LBGs):\\\\
  -- GALEX-selected sample: good match in IR/UV-metallicity,
  morphology, wind properties, $f_{esc}$(Lyc), $v$/$\sigma$
  ($\uparrow$ with increasing $M_{\star}$), SSCs. {\em What is the gas
    mass fraction, $f_{gas}$, in these systems and how does it compare
    with the value in LBGs?}  [T.\ Heckman, R.\ Overzier, T.\ Gon\c
  calves, Poster: C.\ Cortijo Ferrero]
\item
Local analogues to IR-bright galaxies:\\\\
-- Sub-mm galaxies $\sim$ local ULIRGs in terms of $\Sigma$(SFR) [K. Men\'endez Delmestre]\\\\
-- Size difference is likely due to difference in $f_{gas}$: $\sim$0.15 vs 0.45-0.5 at high redshifts
\item
Local analogues to Ly$\alpha$ emitting galaxies:\\\\
-- Dwarf irregulars? [not discussed at this workshop]
\item
  Near-field cosmology via Galactic archeology:\\\\
  -- Detailed 3D kinematics + detailed spectroscopic
  studies of stars in nearest galaxies starting with
  the Milky Way [not discussed at this workshop]
\end{itemize}

\subsubsection{Mass assembly: mergers vs cold gas accretion}

\noindent
Some of the talks at this workshop discussed the overall
importance of mergers in triggering starbursts and galaxy evolution:

\begin{itemize}
\item
Last 6-7 Gyr (z$<$0.8):\\\\
-- GEMS: (1) Mean SFR of ``visibly merging systems'' is only modestly enhanced. (2) Visible mergers contribute less than 30\% of $\rho_{SFR}$ (seems consistent with merger simulations) [S. Jogee].\\\\
-- CDFS + IMAGES: images + 2D kinematics suggest that major mergers contribute to change of Spiral/Peculiar galaxy ratio: 31/52\% to 72/10\% (no change in E/S0) $\rightarrow$ case of disk + disk = disk (high $f_{gas}$ [R. Delgado Serrano, F. Hammer]
\item
z$\gsim$ 1-2:\\\\
-- Sub-mm galaxies: OSIRIS/IFU reveal extended ($\gsim$ 8--12 kpc) H$\alpha$ nebulae without well-ordered rotation $\rightarrow$ indicative of mergers triggering starbursts with 10--100 M$_{\odot}$ yr$^{-1}$ kpc$^{-2}$ [K. Men\'endez Delmestre].\\\\
-- Gravitationally lensed systems: IFU probe $\sim$100 pc scale, reveal compact rotating disk with
$<$Q$>\sim$~0.6 $\rightarrow$ unstable to collapse, high $\Sigma$(SFR) but follows Kennicutt law [M. Swinbank].\\\\
--Spitzer/Herschel: (1) Improved $M_{\star}$, photo-z, contributions from obscured star formation and AGN activity. (2) Showed and quantified downsizing (more massive galaxies form earlier). (3) L(TIR) derived from 24$\mu$m + local templates often overestimates actual L(TIR) in objects at z~$>$1.5 [P.\ P\'erez Gonz\'alez]
\end{itemize}

\subsection{5. Future Prospects: 0-10 years}

\noindent
Future prospects for the next decade can first be divided into
theoretical/numerical challenges and observational requirements and
then into a few obvious subcategories:

\begin{itemize}
\item
Theoretical/numerical challenges:\\
\hspace*{1.0cm}-- Pray that Moore's law continues....\\
\hspace*{1.0cm}-- Higher dynamical range in spatial scales\\
\hspace*{1.0cm}-- More realistic ISM (multi-phase, shocks....)\\
\hspace*{1.0cm}-- Better prescriptions from observers\\
\item
Observational requirements\\
\hspace*{1.0cm}-- Higher sensitivity\\
\hspace*{1.0cm}-- Better spatial resolution\\
\hspace*{1.0cm}-- Broader wavelength range\\
\end{itemize}

\noindent
The remainder of this section discusses observational facilities that will make the task of studying starburst galaxies, near and far, easier. 

\vskip 0.2in


\noindent{\bf -- LATEST FIR Facilities:}

\begin{itemize}
\item
Herschel Space Observatory (2009-2012)\\
\hspace*{1.0cm}-- 3.5 meter telescope\\
\hspace*{1.0cm}-- PACS/SPIRE/HIFI\\
\hspace*{1.0cm}-- $\sim$55 -- 671 $\mu$m\\
\hspace*{1.0cm}-- Resolution: $\sim$5-60''\\
\item 
Planck Surveyor (2009-2012)\\
\hspace*{1.0cm}-- All-sky CMB experiment, but pesky Milky Way in the foreground!\\
\hspace*{1.0cm}-- $\sim$350 -- 10,000 $\mu$m\\
\hspace*{1.0cm}-- Resolution: $\sim$5-33'\\
\item
SOFIA (2010-...)\\
\hspace*{1.0cm}-- Flight light: May 26!\\
\hspace*{1.0cm}-- 2.5 meter telescope\\
\hspace*{1.0cm}-- $\sim$300 -- 1600 $\mu$m\\
\hspace*{1.0cm}-- Resolution: $\sim$0.5-3'\\
\end{itemize}

\noindent{\bf -- FUTURE FIR Facilities:}
\begin{itemize}
\item
ALMA (North America/Europe/Japan)\\
\hspace*{1.0cm}-- Interferometer w/ 66+ antennae (5+1/month)\\
\hspace*{1.0cm}-- $\sim$350 -- 3600 $\mu$m\\
\hspace*{1.0cm}-- Resolution: $\sim$0.005-0.05''\\
\hspace*{1.0cm}-- Early Science: $\sim$mid-2011\\
\hspace*{1.0cm}-- Completion: $\sim$2013\\
\item
Large Millimeter Telescope (Mexico/UMass)\\
\hspace*{1.0cm}-- 50-meter single dish\\
\hspace*{1.0cm}-- $\sim$850 -- 4000 $\mu$m\\
\hspace*{1.0cm}-- Resolution: $\sim$5-20'' and FOV up to $\sim$8'\\
\hspace*{1.0cm}-- Completion: $\sim$2011\\
\item
Cornell-Caltech Atacama Telescope (CCAT)\\
\hspace*{1.0cm}-- 25-meter single dish\\
\hspace*{1.0cm}-- $\sim$350 -- 1400 $\mu$m (perhaps broader range)\\
\hspace*{1.0cm}-- Resolution: $\sim$4-16'' and FOV $\sim$10 - 20'\\
\hspace*{1.0cm}-- Completion: $t_0$ + 6 years, where $t_0$ is starting time\\
\item
SPICA (JAXA/ESA/NASA?;2020?)\\
\hspace*{1.0cm}-- 3.5 meter telescope at $\sim$4.5 K\\
\hspace*{1.0cm}-- SAFARI/BLISS/mu-spec\\
\hspace*{1.0cm}-- Precursor to CALISTO/SAFIR and SPIRIT/SPEC...\\
\end{itemize}

\noindent{\bf -- FUTURE OIR Facilities:}
\begin{itemize}
\item
WISE: Wide-Field Infrared Space Explorer\\
\hspace*{1.0cm}-- 16'' telescope launched Dec 2009\\
\hspace*{1.0cm}-- $\sim$10 month all sky survey\\
\hspace*{1.0cm}-- 3.4, 4.6, 12 and 22 $\mu$m\\
\hspace*{1.0cm}-- Resolution: 6 -- 12''\\
\item
JWST($\sim$2014-2020+)\\
\hspace*{1.0cm}-- $\sim$6.5m telescope\\
\hspace*{1.0cm}-- NIRCam/NIRSpec/TFI/MIRI\\
\hspace*{1.0cm}-- Imaging and spectroscopy\\
\hspace*{1.0cm}-- $\sim$0.6 -- 29 $\mu$m\\
\hspace*{1.0cm}-- Resolution: $\sim$0.06 - 0.6''\\
\item
ELTs: GMT/TMT/E-ELT ($\sim$2016-....)\\
\hspace*{1.0cm}-- Large light buckets: $\sim$400, 600, 1200 m$^{2}$\\
\hspace*{1.0cm}-- MCAO $\rightarrow$ $\sim$0.009 - 0.005 (at 1 $\mu$m)\\
\hspace*{1.0cm}-- OH suppression $\rightarrow$ $\sim$1/20 x bkgd (JH bands)\\
\end{itemize}

\end{document}